\documentclass[letterpaper, 10 pt, conference]{ieeeconf}  
\IEEEoverridecommandlockouts                              
\usepackage{graphicx}
\usepackage{amsmath} 
\usepackage{amssymb}  
\usepackage{float}

\usepackage[ruled,lined]{algorithm2e}
\usepackage[skip=0pt]{caption}
\usepackage[skip=-2pt]{subcaption}
\usepackage{verbatim} 
\usepackage{tcolorbox} 
\newtcolorbox{todobox}{colback=red!5!white,colframe=red!75!black} 
\usepackage{xcolor}
\usepackage{breqn} 
\usepackage{outlines}
\usepackage{multirow}

\title{\LARGE \bf
Fleet Sizing for the Flash Delivery Problem from Multiple Depots a Case Study in Amsterdam
}

\author{Maximilian Kronmueller$^{\star}$, Andres Fielbaum$^{\star}$ and Javier Alonso-Mora$^{\star}$ 
\thanks{$^{\star}$ Department of Cognitive Robotics at the Faculty of Mechanical, Maritime and Materials Engineering, Delft University of Technology}
}

\begin{document}
\maketitle

\thispagestyle{plain}
\pagestyle{plain}

\begin{abstract}
In this paper, we present a novel approach for fleet sizing in the context of flash delivery, a time-sensitive delivery service that requires the fulfilment of customer requests in minutes. Our approach effectively combines individual delivery requests into groups and generates optimized operational plans that can be executed by a single vehicle or autonomous robot. The groups are formed using a modified routing approach for the flash delivery problem. Combining the groups into operational plans is done by solving an integer linear problem.
To evaluate the effectiveness of our approach, we compare it against three alternative methods: fixed vehicle routing, non-pooled deliveries and a strategy encouraging the pooling of requests. The results demonstrate the value of our proposed approach, showcasing its ability to optimize the fleet and improve operational efficiency. Our experimental analysis is based on a real-world dataset provided by a Dutch retailer, allowing us to gain valuable insights into the design of flash delivery operations and to analyze the effect of the maximum allowed delay, the number of stores to pick up goods from and the employed cost functions.
\end{abstract}

\section{INTRODUCTION}
In the ever-evolving landscape of retail and logistics, the prominence of flash deliveries as a powerful business model is evident through the success of young companies like Flink, Getir, and Gorillas. The growing demand for instant gratification and swift order fulfillment has been the driving force behind the surge in popularity for this time-sensitive delivery approach. Flash deliveries provide customers with the convenience of receiving their requests promptly, challenging traditional retailers to adapt and secure their market share in this highly competitive arena. Collaborations between established players in the industry further exemplify the industry's response to this trend.
For instance, in the Netherlands, Albert Heijn partnered with Thuisbezorgd and Deliveroo to provide faster grocery delivery \cite{albert_thuisbezorgd}. Similarly, Cornershop merged with Uber pursuing similar objectives \cite{cornershop_uber}.

These processes are accelerated and challenged further by the rapid progress in autonomous delivery robots and autonomous driving technologies. A notable example is Starship Technologies, which has successfully completed millions of autonomous deliveries using their robot solution \cite{starship_homepage}. This advancement opens up possibilities for operating large fleets at reasonable costs.

In contrast, a potential upside that traditional retailers have is the ban on opening new dark stores, as seen in cities like Amsterdam \cite{darkStore_issue_thesis}. Dark stores serve as dedicated pick-up locations, and the prohibition on their establishment presents an opportunity for brick-and-mortar stores to step in and utilize their existing infrastructure as depots. This allows traditional retailers to leverage their physical presence to support flash delivery operations.

In planning for flash delivery operations, two critical factors come into play: the efficient routing of vehicles or robots and the design of the fleet. The interplay between these factors adds complexity to the overall system. Traditional routing assumes a fixed number of vehicles as input and focuses on optimizing their usage. Fleet sizing involves determining the optimal number of vehicles required based on how they are utilized during service.
To enable fleet sizing with sophisticated routing, we propose a novel method that combines routing optimization and fleet design while considering multiple stores. Taking into account the unique characteristics of multiple stores is particularly important, as it closely resembles the operational setup of traditional retailers. In our study, we utilize real-world data from a Dutch retailer, including information on the locations and number of brick-and-mortar stores and real-life demand patterns. By integrating these aspects, we aim to gain valuable insights into the optimization of flash delivery operations.

The contributions of this work are twofold: First, we propose a novel combination of methods enabling fleet sizing, including vehicle routing for flash delivery operations from multiple stores.
Second, we analyze a real-life scenario of deliveries, emulating the entrance of traditional retailers into the flash delivery market, shading light onto these operations.

\section{RELATED LITERATURE} \label{sec:sota}
This work deals with fleet sizing for the flash delivery problem (FDP), including the pooling of requests, i.e. vehicles can simultaneously carry multiple requests with similar destinations. 
To the best of our knowledge, this work is the first to do so. As such, this related literature section focuses on fleet sizing in general and on routing for the FDP individually.
The related literature predominantly originates from the area of transporting people, such as the dial-a-ride problem and ridesharing. However, there are three key distinctions between these areas and flash delivery logistics with autonomous vehicles or robots. First, in logistics, the pick-up location of a request is ambiguous and needs to be decided on. Second, while minimizing delay is crucial in people transportation, flash delivery prioritizes operational efficiency and resource utilization over delay reduction. Finally, the usage of autonomous vehicles or robots enables continuous operation.

\subsection{Fleet Sizing}
Fleet Sizing generally answers the question, “How many vehicles are required to serve some demand?”. \cite{Andres_scalingEffects} shows that various effects drive these decisions. The existing literature offers two primary categories of approaches: simulation-based and chaining-based methods.
Simulation-based approaches aim to identify optimal fleet designs and sizes by simulating operations with different fleet compositions. For example, in \cite{Fagnant_Kockelman_simulation_austin}, an agent-based micro-simulation model was employed to analyze shared ride services in Austin, Texas. Through cost estimates and simulations with varying fleet sizes, an optimal fleet size was determined using the Golden Section Search method \cite{shao_goldenSearch2008}.
Chaining-based approaches, on the other hand, involve sequencing requests into chains by reallocating vehicles from completed tasks to subsequent ones. The concept of chaining was initially introduced by \cite{Vazifeh_ratti_manhattan} to address the minimum fleet problem for taxi rides in Manhattan. They utilized a shareability graph and applied a maximum matching algorithm to find the minimum fleet.

Building upon chaining, several papers extended the approach to ridesharing applications, enabling vehicles to serve multiple requests. For instance, in \cite{Wallar_vehicleDistAndFS_ICRA}, chains were iteratively formed using an Integer Linear Programming (ILP) solver, progressively extending existing chains by adding new tasks. \cite{Benefits_ridesharing_FS} presented a combined optimization model that integrated pooling and chaining, demonstrating the potential for reduced fleet sizes through pooling. \cite{Wang_FD_ridesharing} proposed a novel order graph capturing complex inter-order shareability and solved a coverage problem over the graph to determine the required fleet sizes. 
The following two works share a similar idea to the approach presented here, which involves initially calculating how requests or passengers can be served together and then applying chaining. In \cite{Qu_HowManyVehicles}, a routing approach and demand forecasting were employed to maximize their proposed utility metric called "demand utility" on shared trips, with chaining based on \cite{Vazifeh_ratti_manhattan}. Additionally, \cite{Balac_Zurich_FD} utilized temporal and spatial aggregation to form trips, formulating fleet sizing as an ILP and solving it as a minimum flow problem.

\subsection{Routing for the Flash Delivery Problem}
The routing aspect of the FDP represents a specialized domain within dynamic vehicle routing problems. While the FDP shares similarities with the Same-Day Delivery Problem, it poses unique challenges by requiring requests to be fulfilled within minutes after being placed rather than by the end of the day.
In the existing literature, only a few works have focused specifically on routing for the FDP, namely \cite{MRS_maxi, kronmueller2023pooled_arxiv}. These studies adopt a rolling horizon approach to address the dynamic nature of the problem by dividing it into multiple snapshot problems. Their methodology involves a two-step process for each snapshot. Firstly, a comprehensive set of potential plans for each vehicle is generated. Subsequently, an assignment problem is solved to determine which plans are executed by which vehicles. These works build the foundation for the routing approach applied in this work. \footnote{Another variation of this approach, generalizing to heterogeneous vehicles, was proposed in \cite{Maxi_ITSC_Heterogeneous}.}

Not focusing on flash delivery, but the instant delivery problem are the works of \cite{zhen_instantDelivery_heterogeneous} and \cite{XUE_accept_schedule_instant}. In \cite{zhen_instantDelivery_heterogeneous}, a column generation approach is used to optimize the assignment of orders to a heterogeneous fleet of vehicles, considering deadlines of up to hours. In \cite{XUE_accept_schedule_instant}, the instant delivery problem with shorter deadlines of 45 minutes is addressed by decomposing it into a series of static problems. Orders are inserted into existing trajectories based on a similarity measure.

\section{PROBLEM FORMULATION}
Intuitively described, the fleet sizing problem poses a problem in which the number of vehicles and their operational plans need to be found to fulfil a given demand. It becomes the fleet sizing problem for the FDP when all requests need to be delivered within the constraints posed by the flash delivery operation. Solutions are optimized based on a given objective.

The inputs are the demand, as a set of requests $\mathcal{R}$ which need to be serviced, the capacity of the vehicles, and a graph $G=(V,E)$ representing the operation environment.
The operational environment is represented as a weighted directed graph denoted as $G=(V,E)$, with vertices $V$ representing different locations $l \in V$ and edges $E$ indicating connections between them. The weight of each edge, denoted as $w(e)$, represents the traversal time. The stores $\mathcal{S}$ form a subset of vertices $V$, where vehicles can pick up goods to fulfil customer requests. All stores have a full stock of goods at all times.

The demand set $\mathcal{R}$ consists of individual customer requests $r=(l^{goal}_r,t_r)$, where $l^{goal}_r \in V$ represents the goal location and $t_r$ is the request placement time. It is important to note that no specific pickup location for each request is specified, as well as no specific set of products, as we assume each request to be unique.
In the FDP, each request must be dropped off within a maximum delay $\rho_r^{max}$, as \cite{MRS_maxi, kronmueller2023pooled_arxiv}. The drop-off delay $\rho_r$ is the difference between the actual drop-off time and the drop-off time if the request was served immediately via the shortest path from the nearest store.
Additionally, we consider fixed times $t^{load}$ to load a request to a vehicle and $t^{deliver}$ to deliver it to the customer.
The assumed capabilities for vehicles are as follows: Each vehicle has a maximum capacity of $\kappa$ and drives along the graph, specifying the needed travelling times.

The objective of the fleet sizing problem is to determine the number of vehicles required and their corresponding operational plans $\omega$. An operational plan consists of an ordered set of locations $l \in V$, where each location is assigned one of the following activities: picking up a request, delivering a request to a customer, or waiting for further instructions. The vehicle follows the shortest path between locations. Consequently, following an operational plan results in the delivery of a set of requests denoted as $o_{\omega}$. Accordingly, the total driving time of a single trajectory $\omega$ is $\phi(\omega)$, and the resulting total delay if following this plan is $\rho(\omega)$. The starting time of an operational plan is $t^{start}_\omega$, and $t^{end}_\omega$ is the ending time, respectively, starting and ending location are $l^{start}_\omega$ and $l^{end}_\omega$.

To execute one operational plan $\omega$, one vehicle is needed. As such, a set of operational plans $\Omega$ can be a solution to the fleet sizing problem if it satisfies certain conditions.
First, to qualify as a solution, together all operational plans $\omega \in \Omega$ must successfully deliver all requests. Thereby, each request must be picked up from a store and delivered to the customer before its specified maximum drop-off time. Second, the capacity of each vehicle must not exceed the maximum capacity constraint.

The evaluation of a solution $\Omega$ is based on the cost function $J(\Omega)$.
The cost function incorporates various factors, including the number of vehicles used (representing fixed capital costs), travel time (representing variable capital costs), and delay costs (representing the quality of service experienced by customers). For each vehicle or executed trajectory in $\Omega$, a fixed capital cost of $M_{fix}$ is incurred. Additionally, the costs of travel time and delay are weighted convexly using a cost weight parameter $\alpha \in [0,1]$.
This results in the cost function as follows: 
\begin{equation} \label{eq:cost_fs_problem}
    J(\Omega) = M_{fix} \cdot |\Omega| + \sum_{\omega \in \Omega}  [(1-\alpha) \cdot \rho(\omega) + \alpha \cdot \phi(\omega)]
\end{equation}

Let $\mho$ be the set that includes all feasible sets of operational plans $\Omega$, representing solutions to the FDP.
Given the set $\mho$, the fleet sizing problem can be formulated as follows:
\begin{equation}
\min_{\Omega \in \mho} J(\Omega)
\end{equation}
Note that constraints are implicit in the set $\mho$.

\section{METHOD} \label{sec:method}
\begin{figure*}
	\centering
	\includegraphics[width=2.\columnwidth]{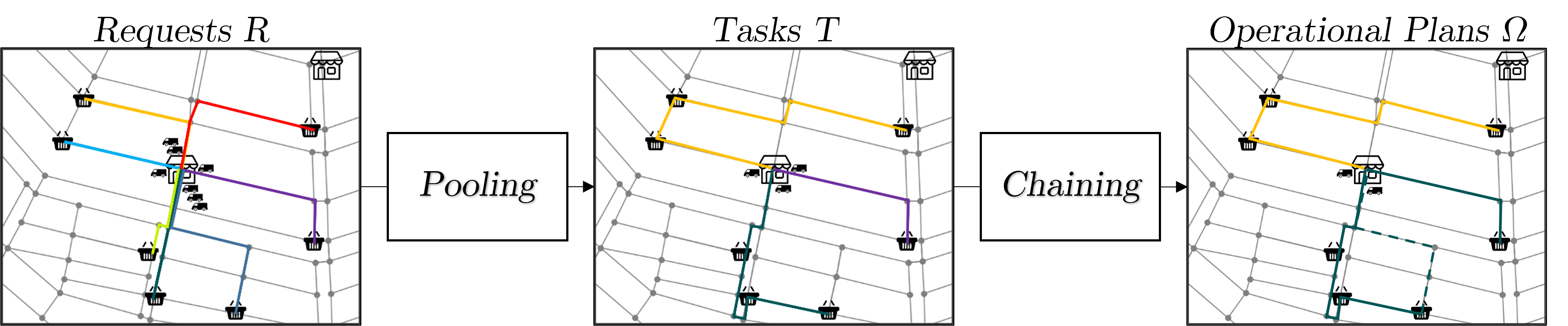}
	\caption{Method Overview: Our method takes a set of requests as input. The first step, pooling, involves grouping the requests into groups that can be efficiently delivered together. These groups are referred to as tasks, which include the corresponding routing optimization. The second step, chaining, focuses on sequencing the tasks to create operational plans, with each plan requiring a single vehicle. To provide visual clarity, different colours are used to represent each vehicle at each step of the process. Dashed lines represent vehicles driving in between tasks (chaining).}
	\label{fig:method_overview}
\end{figure*}

To determine the solution $\Omega_{sol}$, our proposed approach consists of two key steps: pooling and chaining.
Pooling involves the coordinated gathering of multiple requests into groups that can be efficiently delivered together. This step includes optimizing the routing for each group, resulting in small operational plans. For clarity, these small operational plans are not yet the final operational plans spanning the entire operation but rather smaller components. Thus, we refer to these small operational plans as tasks $T$. The notation for tasks is identical to operational plans. For example, a task's starting and ending times are $t^{start}_T$ and $t^{end}_T$, and in the same manner, the starting and ending locations are $l^{start}_T$ and $l^{end}_T$. Intuitively, tasks represent individual units of work that can be performed efficiently by one vehicle.
Chaining combines these small operational plans or tasks $T$ to create final operational plans $\omega$ that cover the entire operation. This is done by assembling the tasks in a sequential manner, considering dependencies and optimizing the overall delivery process.
An overview of this approach is provided in Figure \ref{fig:method_overview}. For a more in-depth understanding of pooling and chaining, please refer to Sections \ref{sec:method_pooling} and \ref{sec:method_chaining}.

\subsection{Pooling} \label{sec:method_pooling}
The pooling step of our method is based on a dynamic routing approach for on-demand last-mile logistics from multiple stores \cite{MRS_maxi, javier_pnas}. However, we adapt this approach to eliminate the requirement of a fixed fleet of vehicles as input, following the methodology proposed in \cite{RSS_cap_multi}. By building upon the principles of \cite{MRS_maxi}, we can ensure that our method generates high-quality routes that satisfy the constraints of the FDP.
To address the dynamic nature of the problem, we employ a rolling horizon approach by dividing the entire operation into multiple snapshot problems.

For each snapshot problem, we first calculate a large set of potential routes for the vehicles. We then select the routes to be executed from this set.
Routes represent vehicle-specific operational plans, considering the vehicle's current state. These routes are designed to be feasible, adhering to the vehicle's capacity constraints and ensuring the timely completion of all assigned requests. Routes overlap multiple snapshot problems and are subject to change.
For algorithmic details on the approach to efficiently calculate the route set, we refer to \cite{MRS_maxi}.  

Each route is assigned a cost to execute it, following Equation \ref{eq:cost_fs_problem}. The set of vehicles to calculate routes for is not fixed in this work but differs for each snapshot problem. In each step, we consider the none idle vehicles of the previous time step and introduce new potential vehicles. We assume that one potential vehicle is available for each request at the closest store to its goal location starting from the request's placement time $t_r$. 

The selection of routes to execute is performed through a coordinated process using an assignment problem, which is formulated and solved as an ILP.
The ILP is a standard formulation to assign routes to vehicles such that all users are served\footnote{Serving all individual requests is always possible due to the possibility of creating new vehicles.} and no vehicle is assigned to more than one route; its explicit formulation can be found in \cite{MRS_maxi,javier_pnas}.
If a new potential vehicle is chosen by the assignment, it is instantiated into the problem and follows the assigned route. Any potential vehicles that are not assigned are disregarded.

Each vehicle follows its assigned route until the next snapshot, at which point the routes of all vehicles are updated and thus can be prolonged.
Once a vehicle completes its assigned route and becomes idle, it is removed from the problem. 
The full route that each vehicle executes, from its creation until it is dropped, constitutes a task $T$. All tasks $T$ are summarized in the set $\mathcal{T}$.

\subsection{Chaining} \label{sec:method_chaining}
The chaining step is employed to combine the tasks $T$ generated by the pooling step into operational plans $\omega$ spanning the entire operation. The objective of chaining is to optimally sequence tasks in a way that allows them to be executed by a single vehicle.

In order for two tasks $T_i$ and $T_j$ to be executed consecutively by a single vehicle, the vehicle must be able to relocate from the end location of task $T_i$, denoted as $l^{end}_{T_i}$, to the start location of the subsequent task $T_j$, denoted as $l^{end}_{T_j}$, and reach it before its designated starting time $t^{start}_{T_j}$. The travel time required to drive from $l^{end}_{T_i}$ to $l^{start}_{T_j}$ is represented as $\tau_{i,j}$.\footnote{During relocation, the vehicle is empty.} Thus, two tasks $T_i$ and $T_j$ can be chained if the following equation is satisfied: $t^{end}_i + \tau_{i,j} \leq t^{start}_j$.
All pairs of tasks $(i,j)$ that fulfil this equation are summarized in the set $\mathcal{X}$.
To coordinate which pairs of tasks from the set $\mathcal{X}$ are actually executed in sequence by one vehicle (chained), an ILP can be formulated and solved \cite{Vazifeh_ratti_manhattan}.
The ILP minimises the overall costs, Equation \ref{eq:cost_fu_chaining}.\footnote{This is equivalent to the overall cost function in Equation \ref{eq:cost_fs_problem}, as the cost to execute tasks can be excluded as it is constant.}
\begin{equation} \label{eq:cost_fu_chaining}
    \min \sum_{i,j \in \mathcal{X}} x_{i,j} \cdot \Big[-M_{fix} + \alpha \cdot \tau_{i,j} \Big]
\end{equation}
Being subject to each task having maximally one preceding and one subsequent task.
Successfully chained tasks form a single operational plan $\omega$ within the solution $\Omega_{sol}$. The size of the solution $|\Omega_{sol}|$ defines the number of required vehicles, as each plan requires one.

\section{DATASET} \label{sec:dataset}
\begin{figure}
     \centering
     \begin{subfigure}[b]{0.45\columnwidth}
         \centering
         \includegraphics[width=\textwidth]{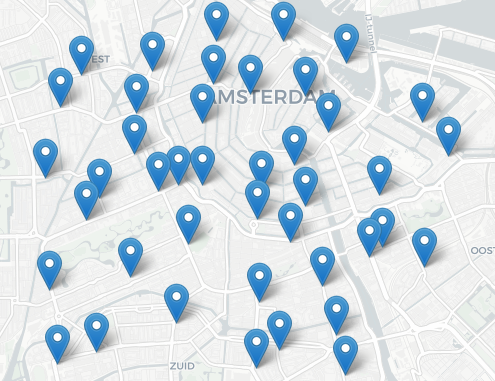}
         \caption{All store locations over a map of Amsterdam.}
         \label{fig:map_full}
     \end{subfigure}
     \hfill
     \begin{subfigure}[b]{0.49\columnwidth}
         \centering
         \includegraphics[width=\textwidth]{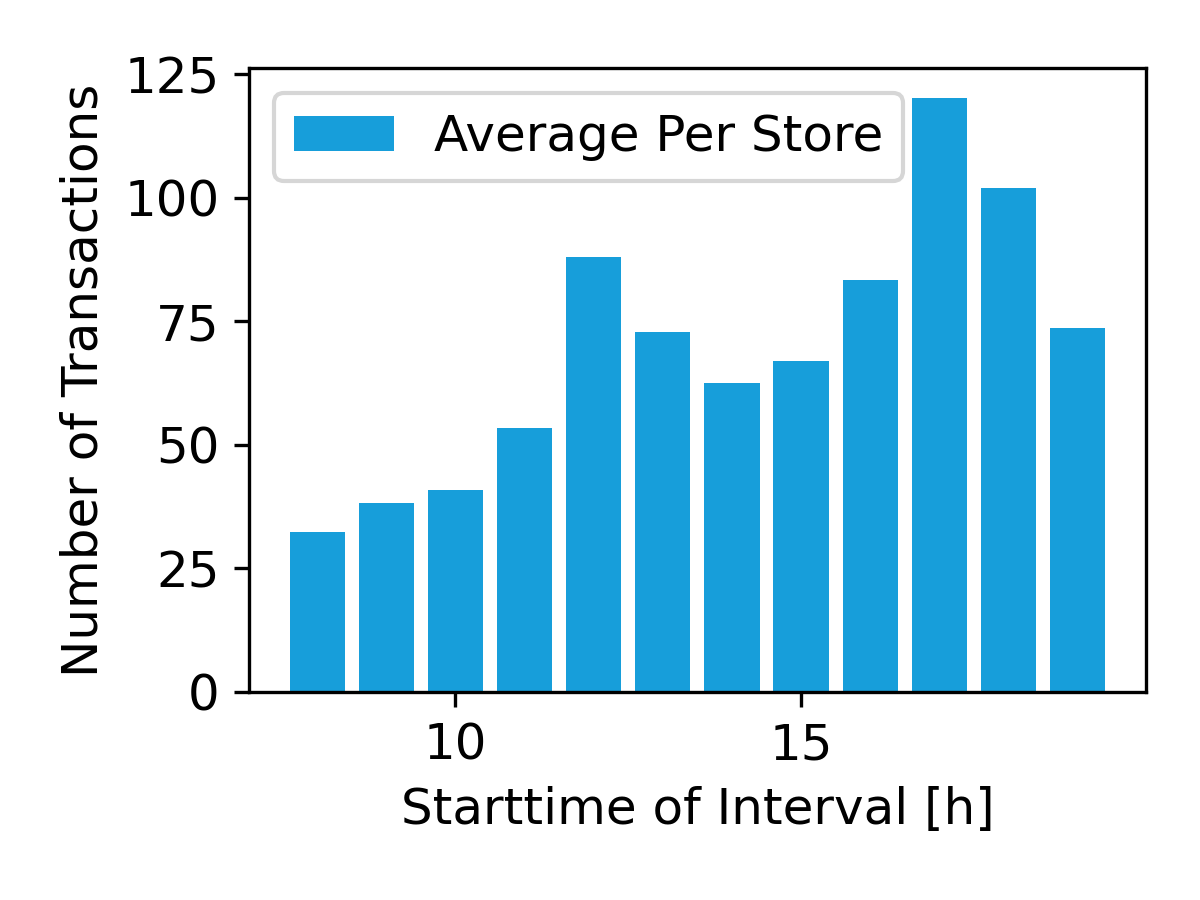}
         \caption{Average number of transactions per store.}
         \label{fig:average_transactions}
    \end{subfigure}
    \vspace{0.35cm}
    \caption{Store distribution and demand data of the used data.}
    \label{fig:data_case_amsterdam}
\end{figure}

The data set this case study is based on describes the shopping behaviour of walk-in customers in regular brick-and-mortar supermarkets in Amsterdam, Netherlands.
The locations of 42 stores belonging to a single retail company in the city center are known and considered as pick-up locations $\mathcal{S}$. Figure \ref{fig:map_full} displays a map of Amsterdam's city center, highlighting the locations of all stores in the dataset. The dataset provides information on the number of transactions per hour for each store, although the exact transaction times are not available. This transaction data is available from 8 a.m. to 8 p.m. Figure \ref{fig:average_transactions} shows the used demand pattern as the average number of transactions for all stores against time. Most notably, clear peaks in demand during noon and the evening are present.
In this study, we simulate a flash delivery operation by modifying the original data. One crucial aspect that undergoes changes is the set of requests $\mathcal{R}$. We presume that people reside in close proximity to the stores they frequent. Specifically, we assume that each person exclusively shops at their nearest store, thereby defining an area $A_s$ associated with each store $s$. This area comprises all vertices $l \in V$ for which store $s$ is the closest one.
We iterate through all stores and time windows to construct the individual requests $r$ within the demand set $\mathcal{R}$. For a given store $s$ and a specific time window $k$ (one hour), the provided data includes the number of transactions conducted at that store. We assume that a constant percentage\footnote{Due to confidentiality reasons, we can not report exact numbers here.} of these transactions will shift from traditional brick-and-mortar stores to the flash delivery service.
For each individual request $r$, we sample the goal location $l^{goal}_r$ from the set of vertices within the corresponding area $A_s$. We assume that customers are uniformly distributed within this area. The request time $t_r$ is also uniformly sampled from the corresponding time window $k$ (one hour).

\section{EXPERIMENTS AND RESULTS} \label{sec:results}
\subsection{Experimental Setup}
Our experiments focus on Amsterdam's city centre, represented by a graph of 2717 vertices and 5632 edges.
However, we reduce the set of stores $\mathcal{S}$. This decision is based on the retailer's reasoning that some stores are too busy to be suitable as pick-up locations. As such, we exclude the busiest half, measured in total number of transactions, of all stores from being pick-up locations. Below we also analyze a scenario using all stores.
The demand, as described above, stays identical, as it is not affected by such strategic decisions.
We assume a loading and service time of 1 minute for each request ($t^{load} = t^{deliver} = 60[sec]$). The maximum delay allowed during pooling is set to 5 minutes ($\rho_r^{max} = 300[sec]$). A snapshot problem is solved every 100 seconds.
Following the logic that delays in on-demand delivery operations are nearly neglectable as long as the request is delivered within the promised time window, we set the cost weight in all functions to $\alpha=1$, fully focusing on total driving time.
In the cost functions, we use a large value of $M_{fix}=2000[sec]$, making the minimization of fleet size the first priority\footnote{$M_{fix}$ larger than the maximal time for relocating, determined by the environment, is sufficient to achieve this effect.}.

Due to the nature of the data used to generate the demand, it exhibits an inherent structure divided into one-hour intervals, clearly seen in Figure \ref{fig:average_transactions}. First, we apply the proposed method to each interval separately. Second, we repeat the chaining step, chaining the obtained operational plans per interval. 
\subsection{Results}
\begin{figure}
	\centering
\includegraphics[width=1.\columnwidth]{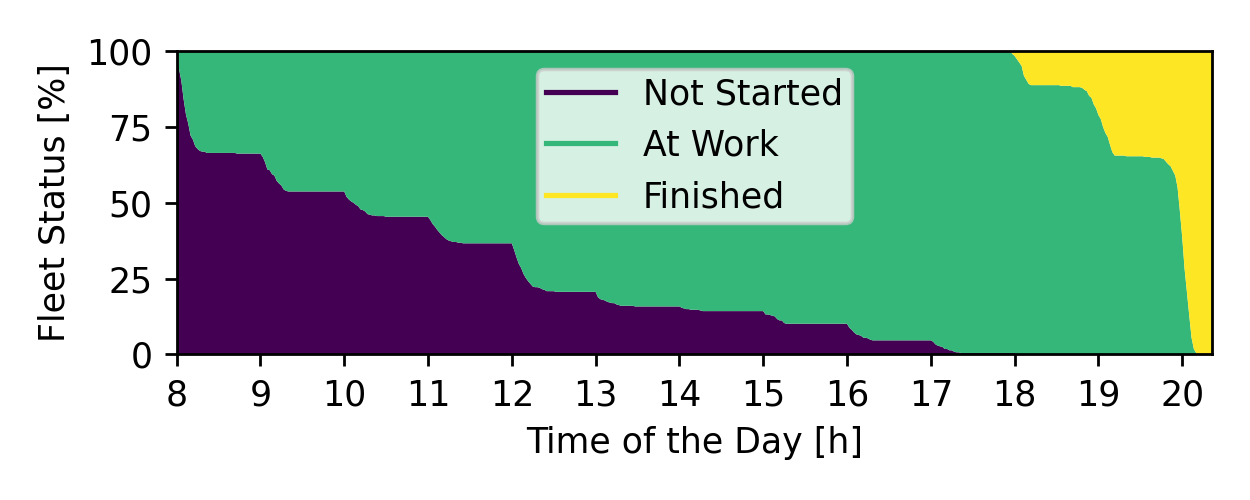}
	\caption{The status of each vehicle throughout the day.}
\label{fig:one_day_fleetStatus}
\end{figure}

\begin{figure}[H]
	\centering
	\includegraphics[width=1.\columnwidth]{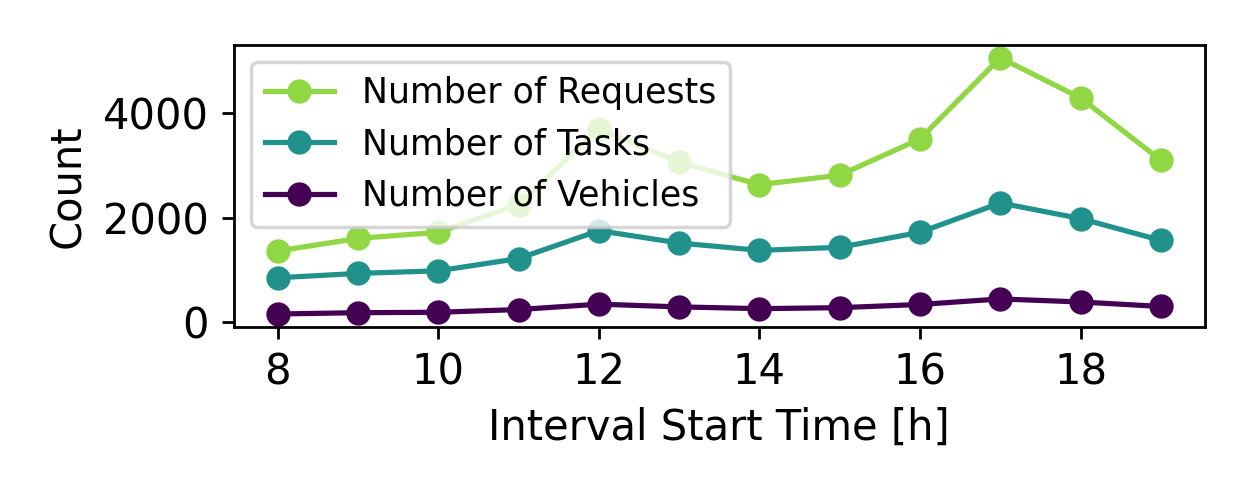}
	\caption{The number of requests, the number of tasks (result of pooling) and the required fleet size for each interval for the entire day are shown.}
	\label{fig:one_day_req_task_fs}
\end{figure}

First, ``How many vehicles are required?''. For the entire day, a total of 459 vehicles are needed. Figure \ref{fig:one_day_fleetStatus} illustrates the status of each vehicle throughout the day. The number of working vehicles (green) increases, i.e. the number of vehicles yet to start (purple) decreases, reaching the highest fleet utilization during the hour with the highest demand. The steps in the graph are due to the hourly segmentation of the demand.
To further understand the results we analyze each hour in more detail.
The number of requests, the number of tasks (the result of pooling) and the required fleet size per hour are shown in Figure \ref{fig:one_day_req_task_fs}.
As the number of initial requests increases in one interval, more tasks are generated, leading to higher fleet sizes. The difference between the number of requests and tasks becomes more significant with higher demand, indicating that it is easier to group and serve requests together when there is a larger volume of demand.
During the peak hour (17:00-18:00), 5054 requests were grouped into 2278 tasks. Each task serves an average of 2.22 requests and takes 445.7 seconds. To handle these tasks, a fleet of 443 vehicles is required, which is slightly less than for the entire day. Each vehicle serves 11.40 requests on average, and an operational plan takes approximately 55 minutes and 33 seconds. This duration is close to spanning the full hour, indicating effective utilization of the vehicles.

\begin{figure}[H]
	\centering
	\includegraphics[width=1.\columnwidth]{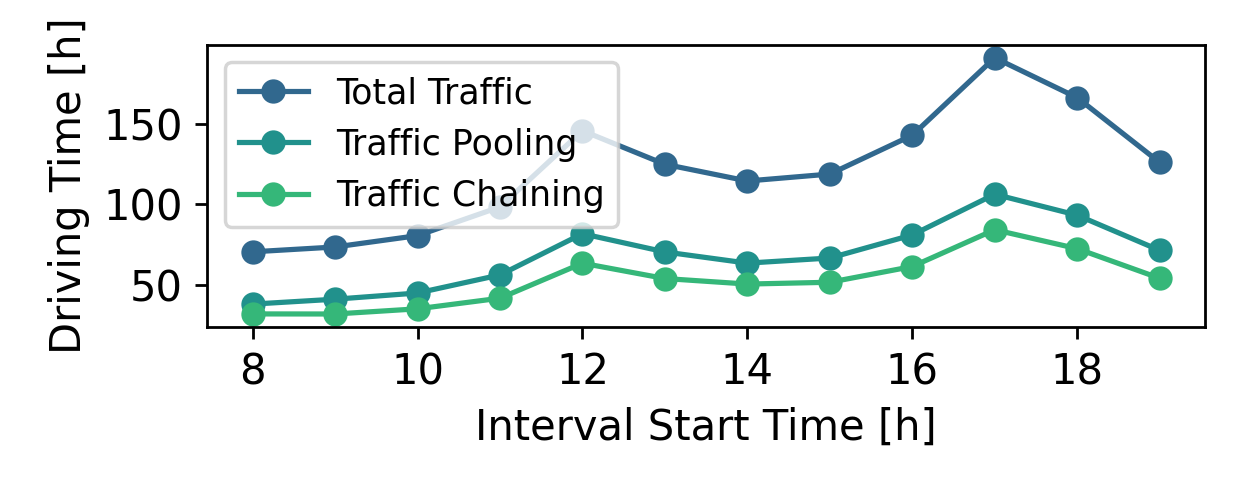}
	\caption{The total traffic and its breakdown into polling traffic and chaining traffic for each interval throughout the day is shown.}
	\label{fig:one_day_traffic}
\end{figure}

Total traffic by all vehicles shows the same correlation with demand (Figure \ref{fig:one_day_traffic}). It is observed that traffic originating from the pooling step (pooling traffic) exceeds traffic originating from the chaining step (chaining traffic). This difference becomes more pronounced with higher demand.
Generally, with higher demand, the average chaining traffic per vehicle decreases.

\begin{figure}[H]
     \centering
     \begin{subfigure}[b]{0.49\columnwidth}
         \centering
         \includegraphics[width=\textwidth]{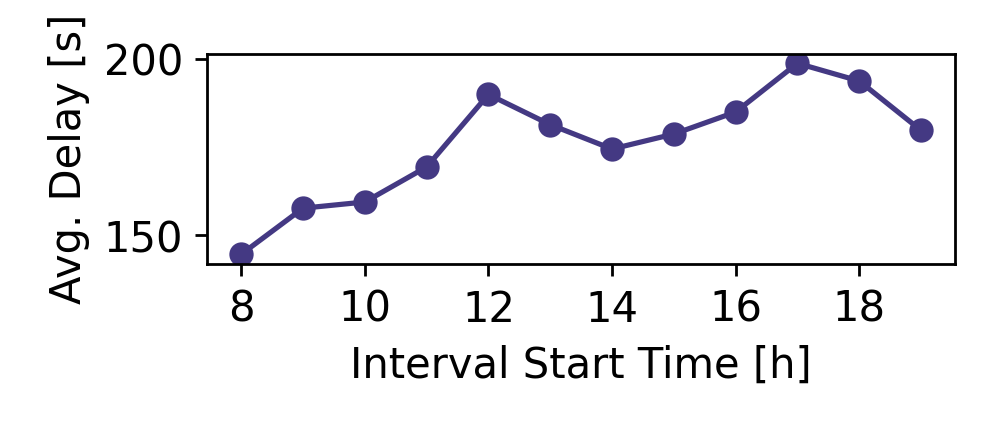}
         \caption{Average delay of intervals.}
         \label{fig:one_day_delay}
     \end{subfigure}
     \hfill
     \begin{subfigure}[b]{0.49\columnwidth}
         \centering
         \includegraphics[width=\textwidth]{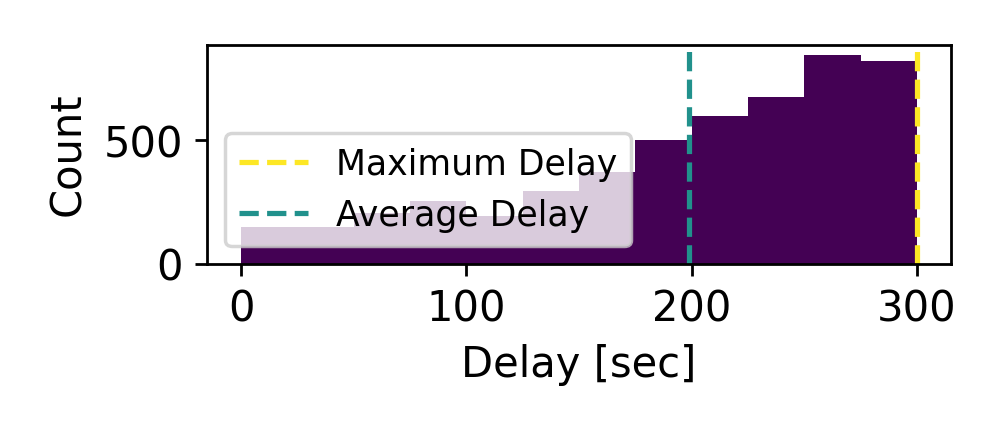}
         \caption{Distribution peak interval.}
         \label{fig:delay_dist_hour_base}
    \end{subfigure}
    \vspace{0.35cm}
    \caption{Two figures showing the average delay of all intervals (a) and the delay distribution of the peak interval (b).}
    \label{fig:delay_main}
\end{figure}

Figure \ref{fig:one_day_delay} presents the average delay per request over the course of the entire day. It differs by about 1 minute between 140 and 200 seconds over the day. About 50 seconds, half of the time step of the pooling algorithm is due to the applied rolling horizon approach.
Additionally, Figure \ref{fig:delay_dist_hour_base} displays the delay distribution for all requests between 17:00 and 18:00. Each request experiences an average delay of 199 seconds. There is a noticeable increase in the number of requests with higher delays approaching the maximum allowed delay of 300 seconds. \footnote{Recall that delay was not considered as part of the cost function. We do so as part of the sensitivity analysis below.}
To conduct a comparative analysis and examine the key parameters, the focus of the study is narrowed down to the peak hour interval, from 17:00 to 18:00. This time period is chosen due to its significance, as it represents the hour with the highest number of transactions throughout the entire day. 

\begin{figure}
	\centering
	\includegraphics[width=1.\columnwidth]{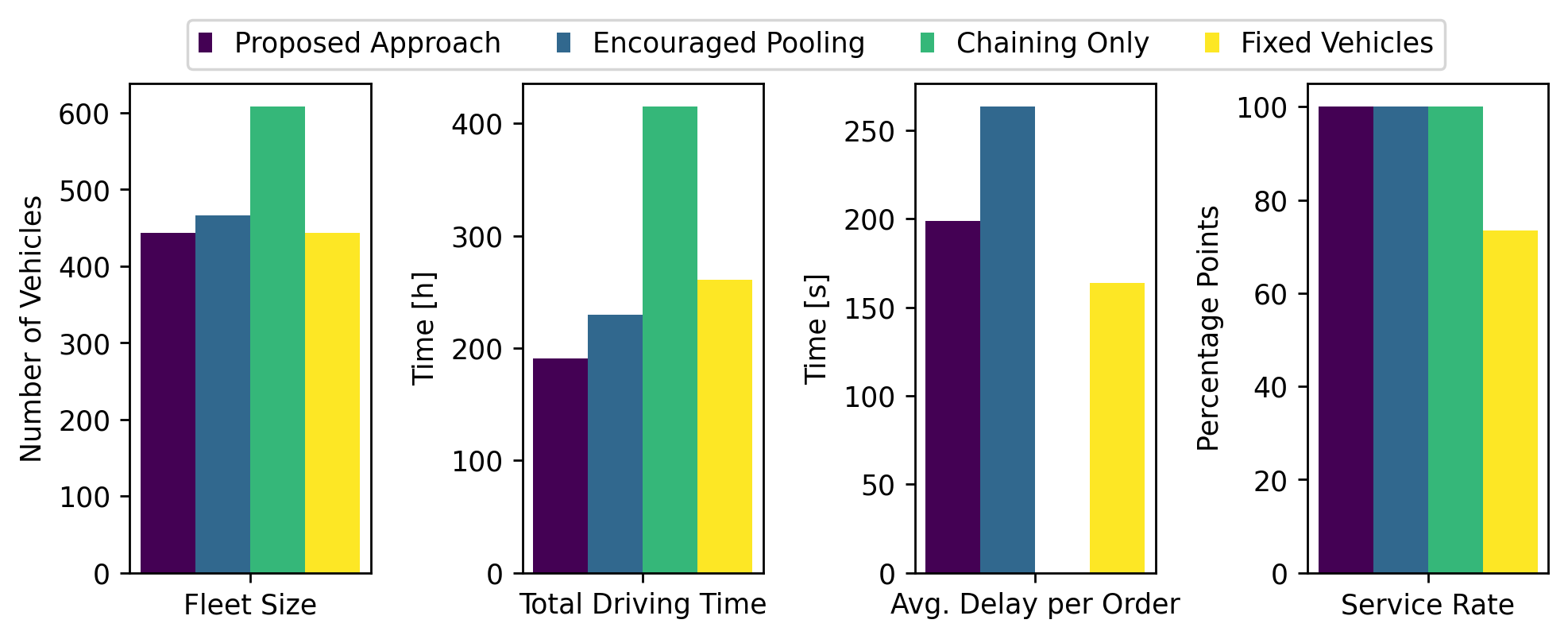}
	\caption{Comparison of the proposed approach to three different strategies based on the main KPIs.}
	\label{fig:experiments_strat_compare_bars}
\end{figure}

To range in the performance of the proposed approach, we compare it against three opposing approaches.
First, `encouraged pooling'', we apply a strategy encouraging pooling to decrease the number of tasks obtained. To do so, we add costs to a route if it uses a new potential vehicle. This extra cost was set to equal 1000\;seconds. 
Second, ``chaining only'', we exclude the pooling step and deliver each request individually. 
Third, ``fixed vehicles'', we use a fixed number of vehicles, equivalent to the results of the proposed approach, and route them as \cite{MRS_maxi} (pooling step).
The comparative results are presented in Figure \ref{fig:experiments_strat_compare_bars}.
In the ``encouraged pooling'' and ``chaining only'' approaches, the fleet sizes increase and higher total driving times compared to the proposed approach are needed. The ``chaining only'' approach has no delay since each request is immediately served with its own vehicle. Service rates are at the enforced 100\% for all three methods (Proposed approach, ``encouraged pooling'' and ``chaining only'').
In contrast, using a ``fixed number of vehicles'' does not enforce the service rate but serves as many requests as possible using the available vehicles. We fixed the fleet size to 443, the same number as for the proposed approach. As a result, around 62.5\% of requests are served, requiring more driving time and a lower average delay. The main reason for this difference is that the vehicles are not rebalanced as effectively as with the chaining step, which is done in hindsight with full information over the full planning horizon.

Last, we study the effect of the maximum allowed delay $\rho_r^{max}$, the number of stores to pick up goods and the cost weight between delay and driving time.
We vary the studied variable exclusively and compare fleet size, traffic and delay.

\begin{figure}
	\centering
	\includegraphics[width=1.\columnwidth]{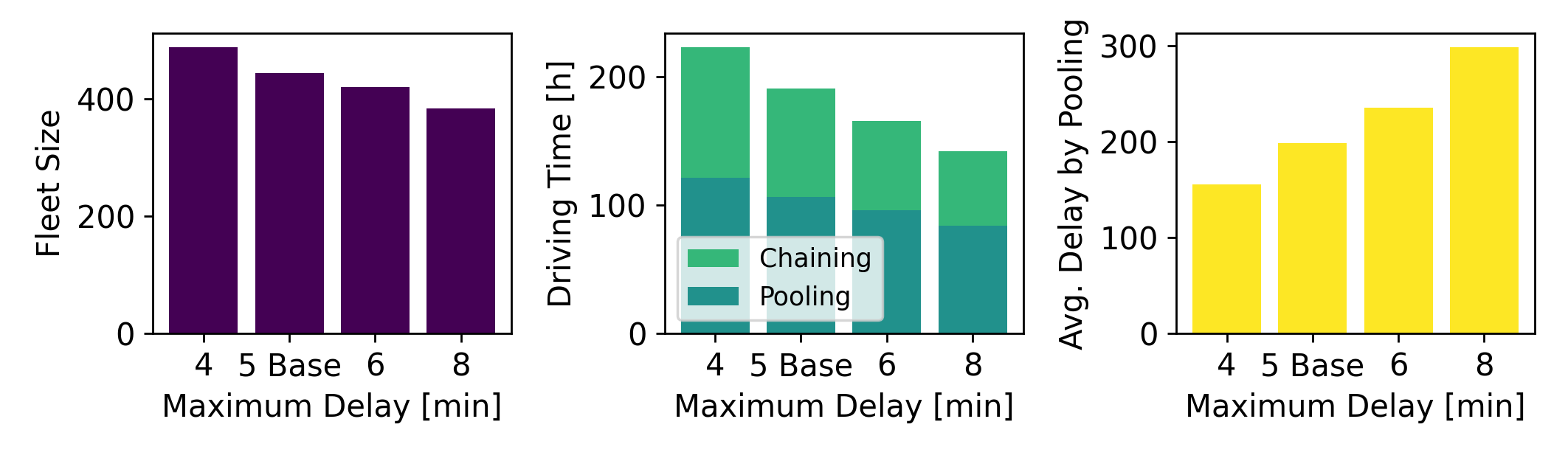}
	\caption{Comparison of fleet size, traffic and average delay for different values of allowed maximum delay.}
	\label{fig:sens_maxDelay}
\end{figure}

\textbf{Delay:} We vary the maximal delay as $\rho^{max} = [4,6,8]$. Results are shown in Figure \ref{fig:sens_maxDelay}.
The higher the maximum allowed delay, the lower are required fleet sizes, accompanied by lower traffic, but at the price of higher values of average delay. This is somehow expected. Most interestingly, are the changes in the split between pooling traffic and chaining traffic. Both decrease with higher maximum delay, but the amount of change in chaining is more, as more requests get served together, which then befits the fleet size. 

\begin{figure}
	\centering
	\includegraphics[width=1.\columnwidth]{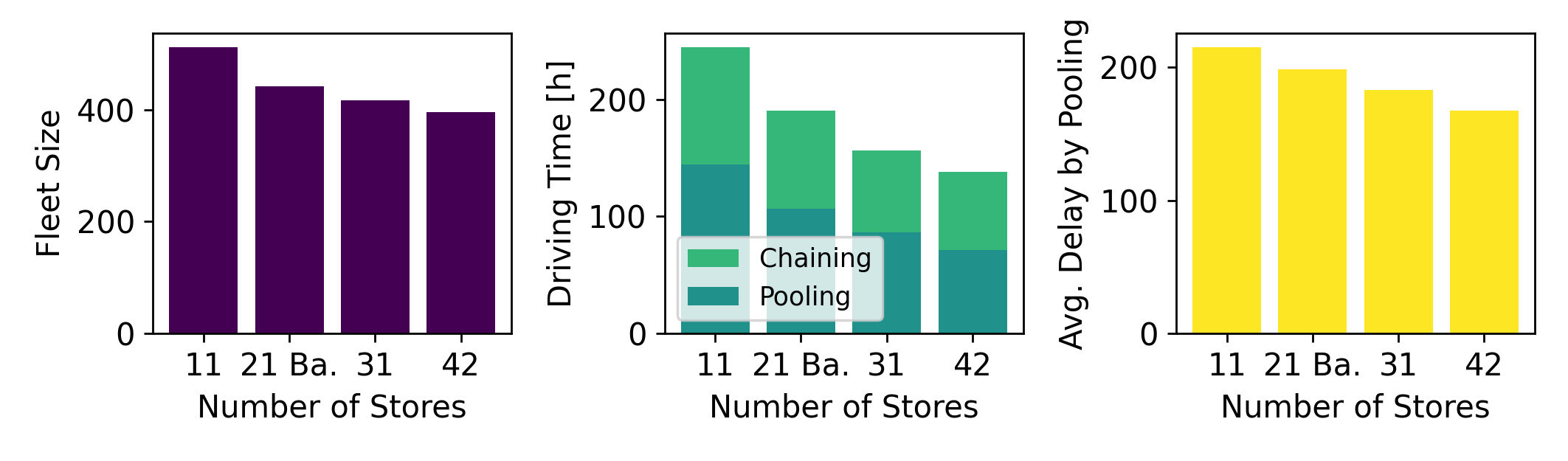}
	\caption{Comparison of fleet size, traffic and average delay for different number of available stores.}
	\label{fig:sens_storesUsed}
\end{figure}

\textbf{Number of Stores:} For the experiments, up to here, the busier half of all stores have not been considered as pick-up locations. Here we compare the influence of the number of used stores on the obtained results. We exclude 10 more and 10 less stores, as well as using all stores of the retailer. Results are visualized in Figure \ref{fig:sens_storesUsed}.
All KPIs improve the more stores are used. For fleet size and traffic the changes get smaller the more stores are used. So the gains of using one additional store when having 11 stores are larger than when already using 41 stores. Changes in delay are more constant. 

\begin{figure}
	\centering
	\includegraphics[width=1.\columnwidth]{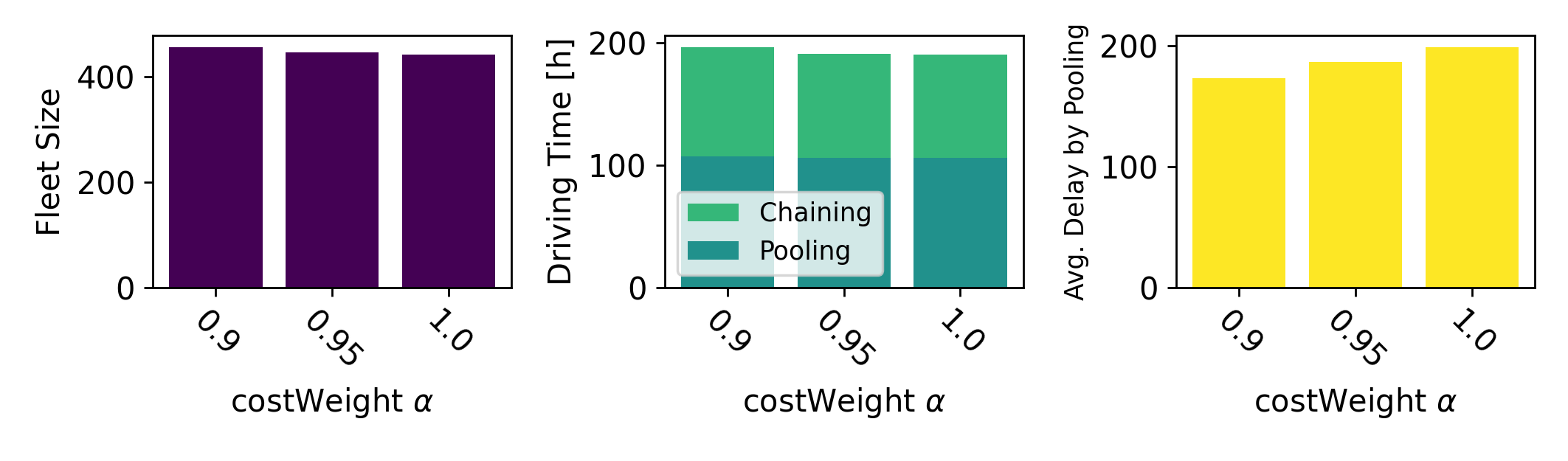}
	\caption{Comparison of fleet size, traffic and average delay for different values of the cost weight $\alpha$.}
	\label{fig:sens_costWeight_alpha}
\end{figure}

\textbf{Cost Weight for Pooling $\alpha$:} The relation between total driving time and delay experienced by customers is captured in the used cost functions. For all experiments so far, we did not consider delay as a cost, here, we do so by varying alpha in $\alpha = [0.9, 0.95]$. Obtained results are illustrated in Figure \ref{fig:sens_costWeight_alpha}. As a direct result average delay decreases, the lower $\alpha$ the more. This comes at the cost of an increased fleet size. Changes in traffic are minor.

\section{CONCLUSION} \label{sec:conclusion}
We presented a novel approach for fleet sizing for the FDP.
The comparison with alternative strategies demonstrates the benefits of our approach, showing that the integration of both pooling and chaining steps leads to improved performance compared to using only one of these strategies.
Furthermore, by utilizing a real-world dataset, we were able to gain valuable insights into the operation of flash delivery services. We explored the effects of store selection, maximum delay, and cost weighting on fleet size, traffic, and delay. These findings provide practical knowledge for designing and managing flash delivery systems in urban environments.
For future work, it is essential to reduce assumptions and incorporate real-life features such as traffic conditions.

\section*{ACKNOWLEDGMENT}
This research was supported by Ahold Delhaize. All content represents the opinion of the author(s), which is not necessarily shared or endorsed by their respective employers and/or sponsors.

\addtolength{\textheight}{-13cm}
\bibliographystyle{IEEEtran}
\bibliography{Library.bib}

\end{document}